\pdfoutput=1
\documentclass[lnbip]{svmultln} 
\usepackage{xspace}
\usepackage{graphicx}
\usepackage[normalem]{ulem} 

\usepackage[usenames]{color}
\usepackage{ifthen}
\usepackage{alltt}
\usepackage{amsmath}
\usepackage{amssymb}
\usepackage{bold-extra}
\usepackage{listings}
\usepackage{needspace}
\usepackage{clrscode} 

\newboolean{showcomments}
\setboolean{showcomments}{false}
\ifthenelse{\boolean{showcomments}}
  {\newcommand{\nb}[2]{
	\fbox{\bfseries\sffamily\scriptsize#1}
    {\sf\small$\blacktriangleright$\textit{#2}$\blacktriangleleft$}
   }
   
  }
  {\newcommand{\nb}[2]{}
   
  }

\definecolor{deiflightblue}{rgb}{0.9,0.9,1}

\usepackage{hyperref}
\hypersetup{
   a4paper,
   colorlinks,
   urlcolor=blue,
   citecolor=blue,
   linkcolor=blue,
   pdftitle = {Collective Behavior},
   pdfauthor = {Adrian Kuhn and David Erni},
}

\RequirePackage{changebar}
\RequirePackage{ulem}
\RequirePackage{color}

\DeclareGraphicsExtensions{.png,.jpg,.pdf,.eps,.gif}

\newcommand{\ie}{\emph{i.e.}\xspace}
\newcommand{\eg}{\emph{e.g.}\xspace}

\newcommand{\OO}{object-oriented\xspace}
\newcommand{\cb}{swarm behavior\xspace}
\newcommand{\Cb}{Swarm behavior\xspace}
\newcommand{\CB}{Swarm Behavior\xspace}
\newcommand{\emphCB}{\emph{Swarm Behavior}\xspace}
\newcommand{\cm}{group method\xspace}
\newcommand{\cms}{group methods\xspace}
\newcommand{\Cms}{Group methods\xspace}
\newcommand{\CMs}{Group Methods\xspace}

\newcommand{\SW}{\textsc{Swarm}\xspace}
\newcommand{\JAG}{\textsc{JavaGroups}\xspace}
\newcommand{\Java}{Java\xspace}
\newcommand{\ttt}[1]{\texttt{#1}}

\newcommand{\needlines}[1]{\Needspace{#1\baselineskip}}

\begin{document}
\mainmatter

\title{Empowering Collections with \CB}

\author{%
	Adrian Kuhn \quad David Erni \quad Marcus Denker 
	}
\authorrunning{%
	A. Kuhn, D. Erni, M.Denker}
\institute{Software Composition Group\\ University of Bern\\
	\href{http://www.iam.unibe.ch/~scg}
	{\texttt{www.iam.unibe.ch/$\sim$scg}}\\
}

\maketitle

\begin{abstract}
Often, when modelling a system there are properties and operations that are related to a group of objects rather than to a single object. In this paper we extend Java with \emphCB, a new composition operator that associates behavior with a collection of instances. The lookup resolution of \cb is based on the element type of a collection and is thus orthogonal to the collection hierarchy. \end{abstract}


\section{Introduction}
\label{sec:intro}

Doug Lea wrote in 1994 \cite{Lea94a}: ``Evidence from over a decade of experience in (non-OO) distributed systems, especially, suggests that groups will become central organizing constructs in the development of large OO systems". Almost 15 years later, many \OO languages still lack dedicated first-class concepts to model the behavior of groups. In existing methodologies and languages, methods are either associated with a class as a whole or with a single instance. However, groups of instances often possess discernible behavior that is not well-captured by current concepts and notations.  

\begin{figure}[h]
 \centering
 \includegraphics[width=.9\linewidth]{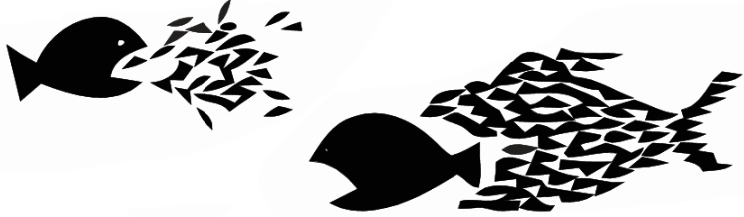}
 \caption{In the \SW game, a collection of fish is under attack by a shark. On the left: the collection of fish cannot defeat the attacking shark. On the right: the collection of fish uses \emphCB to defeat the attacking shark. }
 \label{fig:fishswarm}
\end{figure}

Throughout this paper we use a running example to motivate and explain group behavior. \autoref{fig:fishswarm} illustrates \SW, a small game where creatures fight each other.  Each creature is modeled as an object. Creatures have hitpoints and can attack each other by sending messages.

The user interface of \SW is rather simple. Players select message from pop-up menues to attack opponents. Clicking on a single selected creature opens a pop-up menu with all messages understood by that creature---\emph{but how shall the game deal with multiple selections?}

Modern user interfaces address multiple selection using aggregation and projection. Given a group of selected creatures, the pop-up menu is populated with the intersection of messages that are understood by all creatures in the group (aggregation). When a menu item is selected the according message is sent to all creatures in the group (projection). 

However the behavior achieved through the use of aggregation and projection is limited by the behavior of single elements. Complex behavior cannot be modelled using aggregation and projection alone. \autoref{fig:fishswarm} uses a metaphor to illustrate the difference between simple aggregation and complex group interaction: the collection of fish on the left cannot defeat the attacking shark whereas the collection of fish on the right uses "swarm behavior" to successfully defeat the attacking shark. 

The \SW game might seem an unrealistic example. However, the addressed problem is real in modeling software. There are many realistic applications that would benefit from swarm behavior. The authors of this paper have first come across the missing group-related abstractions when designing the Moose reengineering tool suite, where 20\% of the domain code could be refactored towards \CB \cite{Kuhn07b}. Other applications that would benefit from \cb are file navigators such as the Finder of OSX or any other application that allows multiple selection of domain items. In general any application whose domain model includes grouping could benefit from \CB. Just think how often you have come across classes such as \verb$CustomerList$ or \verb$AccountList$ that extend a general-purpose collection class with behavior specific to their corresponding domain objects.

In this paper we propose \emphCB, a new composition operator that associates behavior with a collection of instances. The lookup of swarm behavior is based on the element type of a collection and is thus orthogonal to the collection hierarchy. 

The contributions of this paper are:
\begin{itemize}
\item We report and analyze the shortcomings of current group-related idioms.
\item We propose \emphCB, a new composition operator that associates the behavior of collection instances with the type of their elements. 
\item We present \JAG \cite{Erni08a}, a research prototype that extends Java's OpenJDK compiler with \CB.
\end{itemize}

\noindent The remainder of this paper is structured as follows: 
\autoref{sec:example} reports on the shortcomings of current group-related idioms.
\autoref{sec:model} presents the model of \CB.
\autoref{sec:java} presents our prototype implementation for Java.
\autoref{sec:javac} provides the details of how we extended the \Java compiler.
\autoref{sec:discuss} discusses issues related to \CB in general and to the \JAG prototype.
\autoref{sec:rel} discusses related work and \autoref{sec:fazit} concludes the paper.

\section{Group-related Idioms in Current Languages}\label{sec:example} 

In this section we report and analyze the shortcomings of current group-related idioms. We present three common idioms that are often used in programming languages (such as \eg Java, Smalltalk, and Ruby) that lack first-class support for group behavior. For each idiom, we implement the \verb$swarmAttack$ method of the \SW example. We examine each implementation and eventually conclude the properties of a possible solution for group-related behavior. We use the following criteria to examine the implementations:
\begin{itemize}
\item In which lexical scope is the \verb$swarmAttack$ method implemented?
\item For which subtype of \verb$Collection$ is the \verb$swarmAttack$ available?
\item Is the availability of \verb$swarmAttack$ restricted to collections of \verb$Fish$ instances?
\item Given a subclass of \verb$Fish$, can it override the behavior of \verb$swarmAttack$?
\end{itemize} 

\subsection{Idiom \#1, Element-specific Subclass of Collection}

This idiom subclasses a general-purpose collection class, limits the type of the collection elements and implements behavior specific for groups of this type. The following source code creates a subclass of \verb$ArrayList$ and implements the \verb$swarmAttack$ method. 

\needlines{10}
\begin{lstlisting}[label=lst:extendcontainer,caption=]
// define element specific subclass
public class FishSwarm extends ArrayList<Fish> {
	public void swarmAttack(Creature creature) {
		int swarmSize = this.size();
		creature.damage(swarmSize * swarmSize);
	}
}
// Create swarm and attack
Creature shark = new Shark();
FishSwarm swarm = new FishSwarm();
for (int n = 0; n < 10; n++) swarm.add(new Fish());
swarm.swarmAttack(shark);
\end{lstlisting}

\emph{Examination:} This solution is fatally tied to a concrete collection subtype. The \cb neither is available for other collection types, such as linked lists or tree sets, nor is it available on array lists created by third-party code.

\subsection{Idiom \#2, Re-open the Collection Hierarchy}

This idiom is only applicable for languages that can re-open classes. The idiom extends the root of the collection hierarchy with a \verb$swarmAttack$ method. Since Java cannot re-open classes, we provide a Ruby example.

\begin{lstlisting}
// re-open array class and append new method
class Array
  def swarm_attack(creature)
    swarm_size = self.size()
    creature.damage( swarm_size * swarm_size )
  end
end
// Create swarm and attack
shark = Shark.new()
swarm = [ Fish.new(), Fish.new(), Fish.new() ]
swarm.swarm_attack(shark)
\end{lstlisting}

\emph{Examination:} Since the \cb is implemented at the root of the collection hierarchy we can possibly invoke the \cb on collections of the wrong element type.

\subsection{Idiom \#3, Static Helper Method}

This idiom uses a static helper method that expects the collection as its parameter. Static helper methods are typically defined in the scope of the element class. The listing below defines a static helper method \verb$swarmAttack$ that expects an object of type \verb$Collection<Fish>$ as first parameter.

\needlines{5}
\begin{lstlisting}
// defines a static helper method
public class Fish extends Creature {
	public static void swarmAttack(Collection<? extends Fish> swarm, Creature creature) {
		int swarmSize = swarm.size();
		creature.damage(swarmSize * swarmSize);
	}
}
// Create swarm and attack
Creature shark = new Shark();
Collection<Fish> swarm = new ArrayList<Fish>();
for (int n = 0; n < 10; n++) swarm.add(new Fish());
Fish.swarmAttack(swarm, shark);
\end{lstlisting}

\emph{Examination:} This solution is flawed by the lack of polymorphism. Given a class \verb$Herring$ that subclasses fish, \cms of herring can neither inherit nor overload \cms of fish. There is not inheritance hierarchy of \cb that would parallel the inheritance hierarchy of the element types.

\subsection{Conclusion of our Analysis}

We conclude our analysis of the three idioms in a table. Beside the three idioms discussed above, the table also includes the \emphCB that we present in the next section as novel composition operator for group-related behavior.

\begin{center}
\begin{tabular}{lcccc}
~ & ~\#1~ & ~\#2~ & ~\#3~ & \CB \\
\hline
Defined in lexical scope of \verb$Fish$.  & -- & -- & yes & yes \\
Available for all subtypes of \verb$Collection$. & -- & yes & yes & yes \\
Limited to collections of \verb$Fish$ instances. & yes & -- & yes & yes \\
Subclasses can override \verb$swarmAttack$. & yes & (yes) & -- & yes \\
\hline
\end{tabular}
\end{center}


\section{The Model of \CB}\label{sec:model}\label{cbmodel}

In this chapter we propose \CB as a new composition operator that aims to avoid the limitations of the idioms in \autoref{sec:example} while combining their strengths. The listing below implements \verb$swarmAttack$ as \CB of \verb$Fish$. An annotation is used to indicate that \verb$swarmAttack$ is a \cm rather than an instance method.

\begin{lstlisting}
public class Fish extends Creature {
	public void attack(Creature creature) {
		creature.damage( 0 );
	}
	@Group
	public void swarmAttack(Creature creature) {
		int swarmSize = this.size();
		creature.damage( swarmSize * swarmSize );
	}	
}
\end{lstlisting}

Please note, that within the body of \verb$swarmAttack$ the pseudovariable \ttt{this} is bound to a collection of fish instances rather than a single fish instance. It is thus save to invokes the collection's \verb$size$ method.

The following listing illustrates a call site of the above \cm. \Cms are invoked on a collection of their class's instances. 
\begin{lstlisting}
// create a swarm and attack
Collection<Fish> swarm = new ArrayList<Fish>();
for (int n = 0; n < 10; n++) swarm.add(new Fish());
Shark shark = new Shark();
swarm.swarmAttack(shark);
\end{lstlisting}

Please recall that  \verb$swarmAttack$ is defined within the lexical scope of \verb$Fish$ but invoked on a collection of fish instances rather than a single instance. We say that \verb$swarm$ is a \emph{container} that contains \emph{elements} of type \verb$Fish$ and method \verb$swarmAttack$ is a \emph{group method} of \verb$Fish$. 

\begin{figure}[t]
\begin{center}
\includegraphics[width=.7\linewidth]{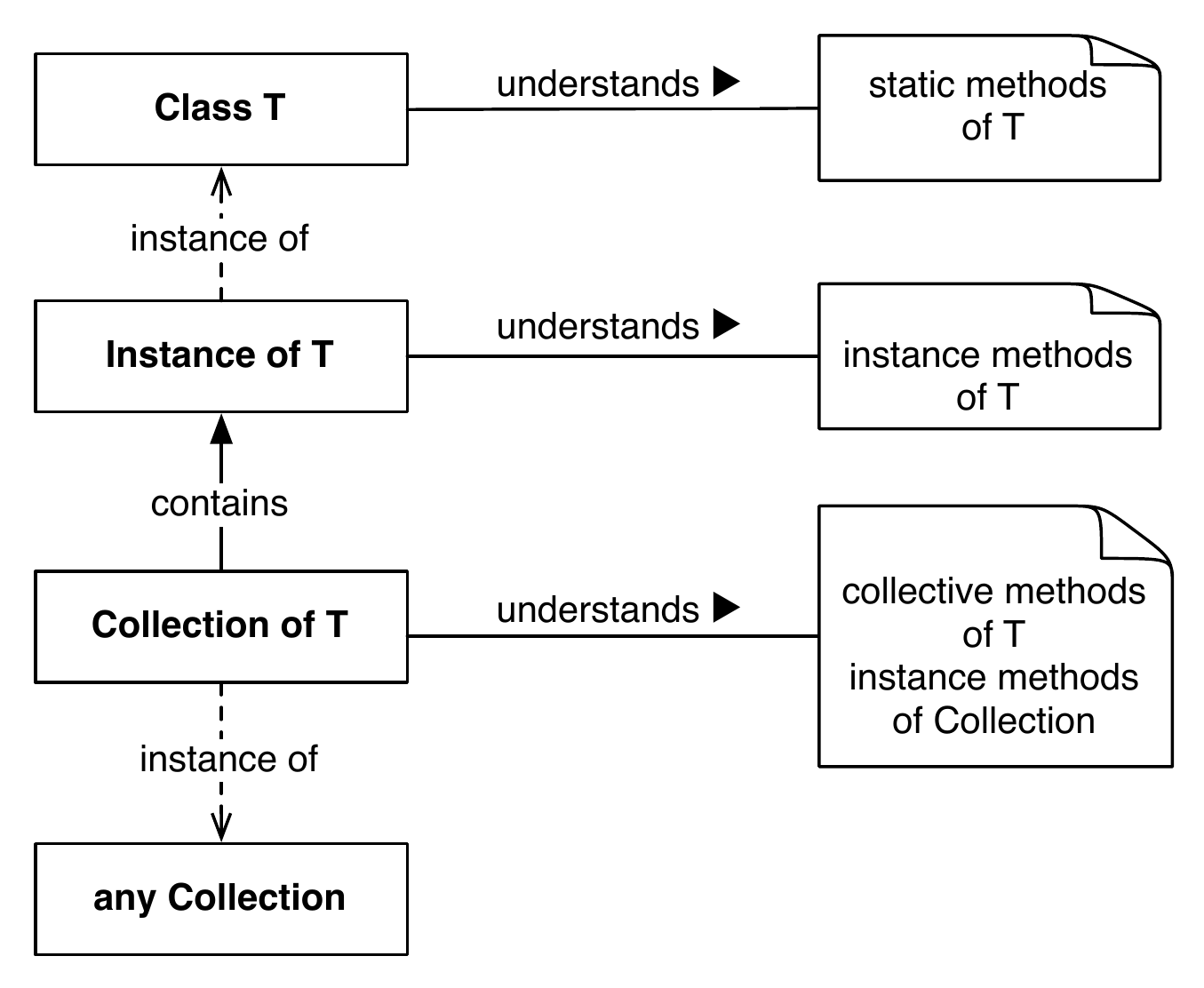}
\caption{The object model of a class \ttt{T} with swarm behavior. \Cms are defined as part of \texttt{T}'s model and invokable on any collection of \texttt{T} instances. Neither are \cms invokable on single instance of type \texttt{T} or nor are they the same as class methods of \texttt{T}.}
\label{fig:diagram}
\end{center}
\end{figure}

\autoref{fig:diagram} illustrates the object model of class \texttt{T} involving \cb. \CB extends the common object model and adds a new option to organize methods. This is achieved by extending the object model with a new kind of methods. In addition to instance and static methods, we allow classes to define group methods. \Cb changes the way method lookup works for any instance of \texttt{Collection} or subclass thereof. The method lookup is extended to take into account both the collection's class hierarchy and the element type. In the same way lookup of \ttt{this} within the lexical scope of \cms is changed.

The procedure $\proc{Swarm-Lookup}(O,S)$ defined the lookup of \emphCB as follows. The procedure has two input parameters, a receiver $O$ and a method selector $S$, and either returns a method $m$ or fails.

\begin{codebox} 
\Procname{$\proc{Swarm-Lookup}(O,S)$}
\li \Comment Lookup of instance methods.
\li $C_\mathcal{I} \gets \id{class}[O]$
\li \While $C_\mathcal{I} \neq \bot$
\li \Do $\mathcal{M}_\mathcal{I} \gets \id{instance-methods}[C_\mathcal{I}]$
\li \If $M_\mathcal{I} \in \mathcal{M}_\mathcal{I}$ and $\id{selector}[M_\mathcal{I}] = S$
\li \Do \Return $M_\mathcal{I}$ \End
\li $C_\mathcal{I} \gets \id{superclass}[C_\mathcal{I}]$
\End
\li \Comment Fail unless receiver is a collection.
\li \If $\lnot(O \id{instance-of} \mathtt{Collection})$ \Return FAIL
\li \Comment Lookup of group methods.
\li $\mathcal{C}_\mathcal{E} \gets \left\{ C : C = \id{class}[E] \land E \in \id{elements}[O] \right\}$
\li $C_\mathcal{S} \gets \proc{Least-Upper-Bound}(\mathcal{C}_\mathcal{E})$
\li \While $C_\mathcal{S} \neq \bot$
\li \Do $\mathcal{M}_\mathcal{S} \gets \id{group-methods}[C_\mathcal{S}]$
\li \If $M_\mathcal{S} \in \mathcal{M}_\mathcal{S}$ and $\id{selector}[M_\mathcal{S}] = S$ 
\li \Do \Return $M_\mathcal{S}$ \End
\li $C_\mathcal{S} \gets \id{superclass}[C_\mathcal{S}]$
\End
\li \Return FAIL
\end{codebox} 

\proc{Swarm-Lookup} works as follows: Lines 2--7 perform an instance lookup in the same way as would be done without \cb. The instance lookup walks up the superclass chain of the receiver's class and returns if a method that matches the method selector $S$ is found. Line 9 stops the lookup unless the receiver $O$ is an instance of \verb$Collection$. Lines 11--12 get the element type of the receiving collection $O$. The element type of a collection is defined as the most specific superclass of all elements in the collection. Lines 13--17 eventually perform the lookup of group methods. Group lookup works the same as instance lookup but differs twofold 1) group lookup starts at the element type $C_\mathcal{S}$ rather than at the receivers class, and 2) group lookup inspects the \emph{group} methods of class rather than the instance methods.

Since by this definition, the \cms of a collection are hidden by its instance methods, an escape is  required to invoke shadowed \cms. Imagine that the implementation of \verb$swarmAttack$ would invoked \verb$Fish.this.size()$ instead of \verb$this.size()$, in this case, the lookup of \verb$size$ would skip the instance lookup and directly start the group lookup with \verb$Fish$ as group class. 

The $\proc{Least-Upper-Bound}$ procedure is defined as follows. Given a collection~$\mathcal{C}$ of classes, the procedure returns the most specific superclass of the input classes. (It is assumed that all classes have one root superclass in common).

\begin{codebox} 
\Procname{$\proc{Least-Upper-Bound}(\mathcal{C})$}
\li $C_0 \gets$ any element of $\mathcal{C}$
\li \For each $C \in \mathcal{C}$
\li \Do \While $C_0 \triangleleft C$
\li \Do $C_0 \gets \id{superclass}[C_0]$
\End
\End
\li \Return $C_0$
\end{codebox}

\section{Extending \Java with \CB}\label{collectivebehavior}\label{sec:java} 

In this section we present \JAG  \cite{Erni08a} a proof-of-concept prototype of \cb. Our implementation extends the Java compiler\footnote{The Java compiler is open source and available as part of the OpenJDK project: \url{http://www.openjdk.org/groups/compiler}.} with support for \cb. In this section we discuss how \JAG modifies the \Java compiler and how the group lookup is achieved. \JAG takes a lightweight approach. We hook into the Java compiler and apply source code transformation such that swarm-enabled source code, which may contain semantically invalid constructs, is transformed to valid java code.

\subsection{Example of \JAG Transformations}

Let's consider the example given earlier in \autoref{sec:model}, that invokes a \cm on a collection of fish. This is invalid \Java code as the \ttt{Collection} does not implement the called method \ttt{swarmAttack}, hence \JAG must transform this code as follows: 

\begin{lstlisting}
// before transformation
Collection<Fish> swarm = new ArrayList<Fish>();
Shark shark = new Shark();
swarm.swarmAttack(shark);

// after transformation
Collection<Fish> swarm = new ArrayList<Fish>();
Shark shark = new Shark();
new Fish.Fish$Group(swarm).swarmAttack(shark);
\end{lstlisting}

The transformed code differs only in the last line. Instead of calling \ttt{swarmAttack} directly on \ttt{swarm}, the swarm collection is wrapped in an element specific subclass of \ttt{Collection} that implements the \cm. The transformed code uses a variation of the ``Element-specific Subclass of Collection'' idiom given in \autoref{sec:example}. We avoid the subclassing limitation of this idiom by creating a proxy that wraps a collection instance. This wrapper class has been generated by \JAG in a previous compiler phase and is described in the following.

In the following listing the class \ttt{Fish} is given, it implements \ttt{swarmAttack} as a \cm. \JAG transforms this code as follows: 

\begin{lstlisting}
// before transformation
public class Fish extends Creature { 	
	@Group public void swarmAttack(Creature creature) { 
		int swarmSize = this.size(); 
		creature.damage( swarmSize * swarmSize ); 
	} 
} 
// after transformation
public class Fish extends Creature  {
	public static class Fish$Group <E extends Fish> implements Collection<E> {
        
		protected Collection<E> delegate;
		public  Fish$Group(Collection<E> delegate){
			this.delegate = delegate;
		}
		public void swarmAttack(Creature creature) {
			int swarmSize = this.size();
			creature.damage( swarmSize * swarmSize );
		}
		// delegation	
		public boolean add(E param0){
			return this.delegate.add(param0);
		}
		// remaining delegation methods
		...
	}
}   
\end{lstlisting}

The transformation moves all methods of a class that are tagged with \ttt{@Group} into a separate class called \ttt{Fish\$Group}. (Out of convenience, \ie namespacing and ease of compiler extension, we decided to use a static inner class. Static inner classes are, despite their name, not inner classes but rather nested top level classes.) \ttt{Fish\$Group} is a proxy that wraps a collection instance. \ttt{Fish\$Group} implements all \cms of \verb$Fish$ as well as the complete interface of \verb$Collection$. The implementation of the interface simply delegates any calls to the enclosed collection instance.

Please  note that, even though \cb rebinds \verb$this$ within \cms, the transformer does not alter any \verb$this$ statement. The \verb$this$ statements are rebound from the fish class to the wrapper class as a side effect of being moved from fish's lexical scope to the scope of the wrapper class.

At this point, the transformation is complete since the example does not make use of inheritance. Inherited \cms require additional transformation which are not discussed here due to space restrictions. For more details, please refer to the \JAG documentation  \cite{Erni08a}. 

\subsection{Inheritance among \CMs}

\begin{figure}[t]
 \centering
 \includegraphics[width=.8\linewidth]{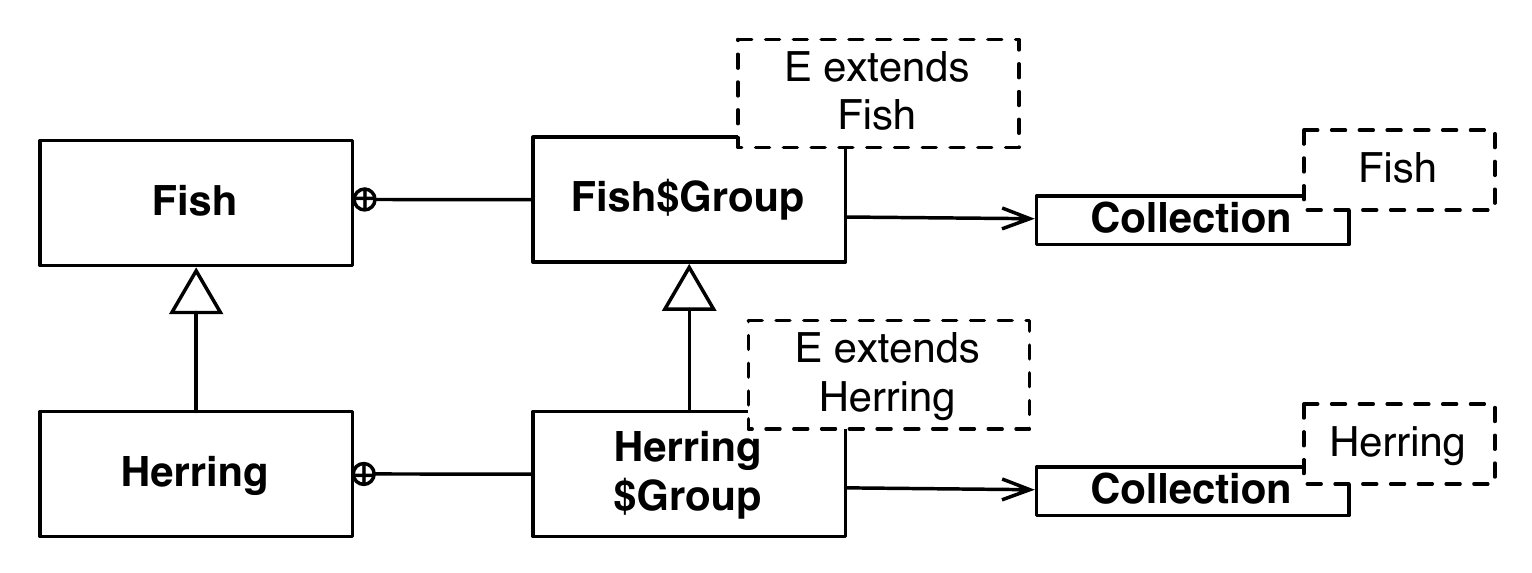}
 \caption[Generating Wrapper Classes as Inner Class including Inheritance and Delegation]{Given two example classes \ttt{Fish} and \ttt{Herring} extending \ttt{Fish}. For both classes a wrapper class is generated and parametrized with its corresponding element type. \ttt{Herring}'s wrapper class extends \ttt{Fish}'s wrapper class just like \ttt{Herring} extends \ttt{Fish}. Both wrapper classes delegate to a parametrized \ttt{Collection}.}
 \label{fig:collectiveclasseswrapper}
\end{figure}

Inheritance among wrapper classes is modeled by paralleling the inheritance hierarchy of element classes amonth the synthetic wrapper classes. \autoref{fig:collectiveclasseswrapper} shows two example classes \ttt{Fish} and \ttt{Herring}, that extends \ttt{Fish}. Both classes have a corresponding wrapper class that is parametrized with the corresponding element type. The type parameter of the wrapper class is set to \ttt{E extends T}. \ttt{Herrings}'s wrapper class extends \ttt{Fish}'s wrapper class because \ttt{Herring} extends \ttt{Fish}, and both of them define \cms. Both wrapper classes delegate to \ttt{Collections} that contain instances of their corresponding elements.

\begin{figure}[t]
 \centering
 \includegraphics[width=\linewidth]{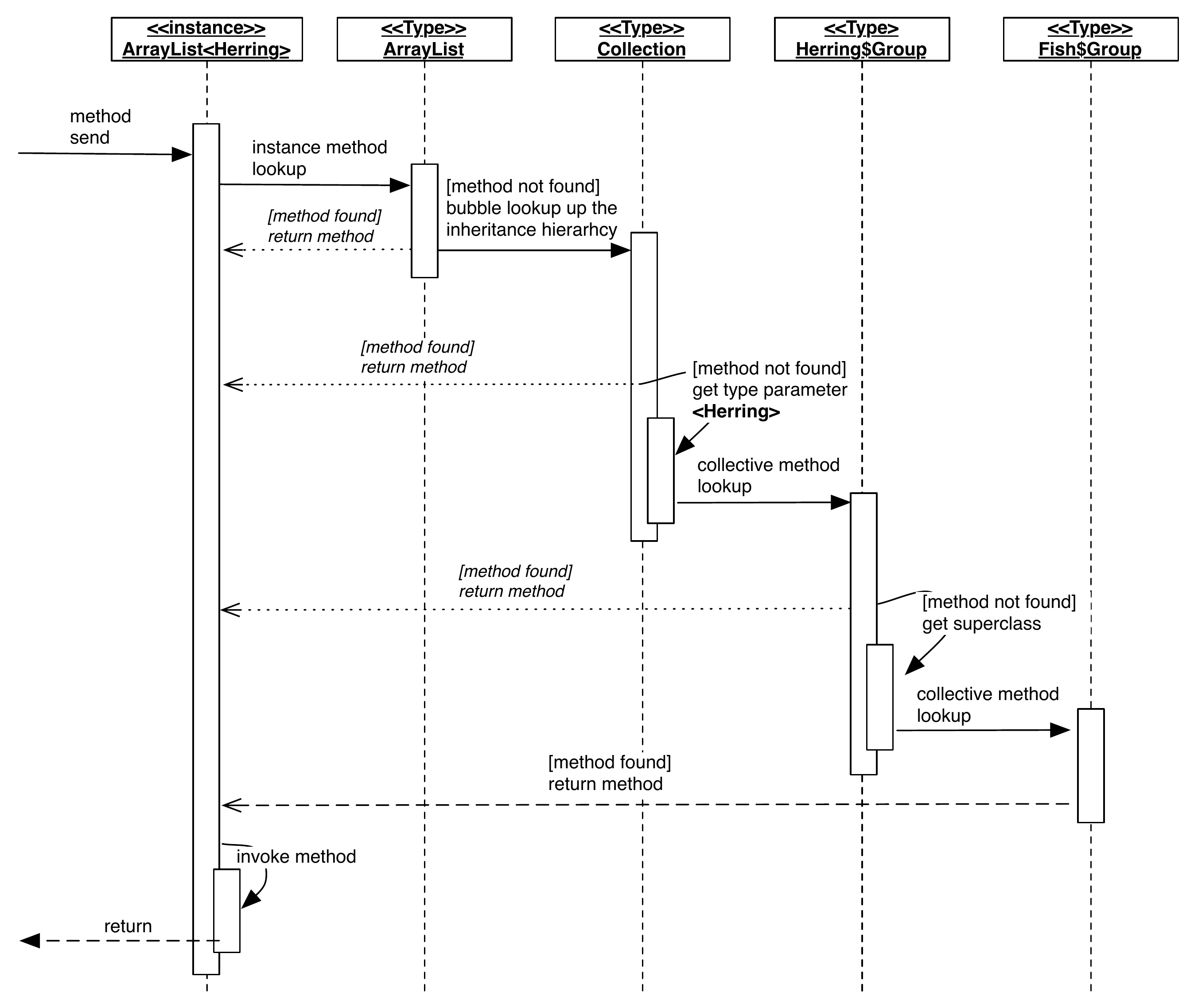}
 \caption[Example Lookup for Collective Methods]{
  Example method invocation on a Collection of Herrings. Let the selected method be a \cm defined on Fish. After the common lookup failed we start the lookup for \cms in Herring. Finally the \cm is found on Fish, returned, and invoked.\footnotemark}
 \label{fig:collectiveclasses}
\end{figure}

\autoref{fig:collectiveclasses} illustrates a sequence diagram of \cm lookup on an instance of \ttt{Container}. Assume we call a method on a \ttt{Collection} of \ttt{Herrings}. The selected method is a \cm defined for a group of \ttt{Fish}, and hence it is available for a group of \ttt{Herrings}, too. First, we check if the method is a method available for instances of \ttt{Collection}. This is not the case, so we start the lookup for \cms in \ttt{Herring} as the container (in this case \ttt{Collection}) is instantiated with  the type parameter \ttt{Herring}. If the \cm is not found for \ttt{Herring}, continue with \ttt{Herring's} supertype, in this case \ttt{Fish}. In that example, the \cm is found in \ttt{Fish's} type and hence lookup is successful. The dotted lines indicate where the lookup process can stop if a method is found. The dashed lines indicate where the lookup stops in this concrete example. 

\section{Modification to the Java Compiler}\label{javacextension}\label{sec:javac}

\footnotetext{\autoref{fig:collectiveclasses} makes use of combined fragments (an UML 2.0 extension), conditional flow of sequence is indicated using dotted lines with conditions given in brackets.}

This section provides the details of how we extended the \Java compiler with \JAG. Readers not familiar with the Java compiler may decide to skip this section. 

\begin{figure}[b]
 \centering
 \includegraphics[width=\linewidth]{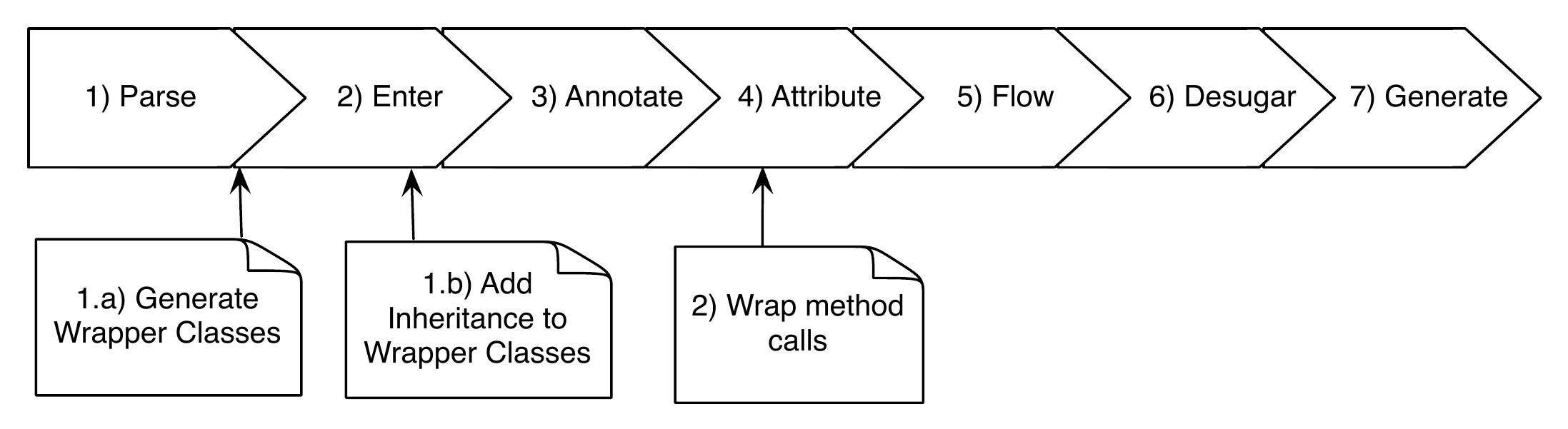}
 \caption[Modified \Java Compiler passes]{Before the Compiler starts the Enter pass, \JAG generates the wrapper classes. Later on during enter, \JAG adds inheritance information to the previously generated wrapper classes. During attribution, \JAG wraps the method calls to \cms into the wrapper classes.}
 \label{fig:jagjavacpasses}
\end{figure}

We start with a brief overview the Java compiler first and give an enumeration of its main compiler phases. \autoref{fig:jagjavacpasses} illustrates the passes that are performed and highlights the points where \JAG differs. The OpenJDK \Java compiler passes these phases 

\begin{enumerate}
 \item \textbf{parse:} Reads a set of *.java source files and maps the resulting token sequence into Abstract Syntax Tree (AST) Nodes.
 \item \textbf{enter:} Enters symbols for the definitions into the symbol table.
 \item \textbf{process annotations:}  If requested, processes annotations found in the specifed compilation units.
 \item \textbf{attribute:} Attributes the syntax trees. This step includes name resolution, type checking and constant folding.
 \item \textbf{flow:} Performs dataflow analysis on the trees from the previous step. This includes checks for assignments and reachability.
 \item \textbf{desugar:} Rewrites the AST and translates away some syntactic sugar.
 \item \textbf{generate:} Generates source files or class files.
\end{enumerate}

\subsection{Transformations to Generate Synthetic Wrapper Classes}

Most of the AST transformation is done during the compiler's Enter pass. 
In step 1a), \JAG transforms the syntax tree by adding the wrapper class as inner class of its corresponding element. In step 1b) \JAG scans the generated classes again to add inheritance information to them.

\subsection{Generate Wrapper Classes}
Let $\mathcal{C}$ be any public toplevel class that gets compiled. We define: 

$$\mathcal{M(C)=\{ }m \mid m \mbox{ is a method of } \mathcal{C\}}$$

$$\mathcal{G(C)=\{ }m \mid m \in \mathcal{M(C) } \wedge m \mbox{ is annotated with @Group }\mathcal{\}}$$

Before the compiler starts the Enter phase, \JAG performs the following transformations for each class $\mathcal{C}$.

\begin{itemize}
\item { \textbf{add an inner class: }
For each class $\mathcal{C}$ where $\mathcal{G(C)} \neq \emptyset$ a wrapper class $\mathcal{C}_\mathrm{group}$ is generated.  This class is added to $\mathcal{C}$'s definitions as inner class. Its name is composed of the string "Group" followed by a dollar sign "\$" followed by the class name of the associated toplevel class. The resulting classname for that inner class $\mathcal{C}_\mathrm{group}$ is $\mathcal{C}$\$Group. 
}
\item { \textbf{creating a constructor:}
Generate a constructor $\mathcal{K}$ for $\mathcal{C}_\mathrm{group}$. As described in \autoref{collectivebehavior} the constructor takes a Collection$<$$\mathcal{C}$$>$ as parameter. This parameter is later used to refer to the original collection for delegation.
}
\item { \textbf{moving \cms:}
Now, each method in  $\mathcal{G(C)}$ is moved to $\mathcal{C}_\mathrm{group}$. This happens by removing it from the definitions of $\mathcal{C}$, and appending it to the definition of $\mathcal{C}_\mathrm{group}$.
Furthermore, the AST of those methods is transformed to use generics. This means that every occurrence of $\mathcal{C}$ as identifier is replaced by the name of the type parameter used for generics, \ie \ttt{E}.
}
\item { \textbf{generating methods from the collection interface:}
The final step for this pass is to generate the methods needed for delegation. Let $\mathcal{F}$ be the containers interface. In our implementation $\mathcal{F}$ is equal to the interface Collection. Furthermore, let 
$\mathcal{I(F)=\{ }m \mid m \mbox{ is defined by }\mathcal{F\}}$.
Now the set

$$\mathcal{D(}m\mathcal{)=\{ }\mbox{m delegates to the Collection passed to }\mathcal{K\}}$$
$$\mathcal{M(F)=\{ }m \mid m \in \mathcal{I(F) } \wedge \mathcal{D(}m\mathcal{)\}}$$

is generated and added to the definitions of $\mathcal{C}_\mathrm{group}$.
}
\end{itemize}

Let $\mathcal{M^{*}(X)}$ be the methods implemented by a class $\mathcal{X}$ after these first translations. Now the methods are distributed as follows: 

$$\mathcal{M^{*}(C)} = \mathcal{M(C)} \setminus  \mathcal{G(C)}$$

$$\mathcal{M^{*}(C}_{group}\mathcal{)} = \mathcal{G(C)} \cup \mathcal{M(D)}$$

\subsection{Enable Inheritance among Synthetic Wrapper Classes}

The second translation step related to \JAG starts as soon as the first step from the original \Java compiler has completed. This includes a detection if the elements that hold wrapper classes reside in an inheritance hierarchy, and if the elements superclasses include \cms, too.
After that original compiler step, all information needed for an inheritance scan is present. This means that the information is accessible by the \ttt{Classsymbols} that are generated during \ttt{ClassEnter} and stored in the appropriate classes. Using those \ttt{Classsymbols} \JAG scans for inheritance recursively, starting from the superclass of the element that contains a \cm. If that superclass contains a suitable wrapper class, it stops the scan and lets the child's wrapper class inherit from that class. Otherwise \JAG continues with the superclasses' superclass until either a wrapper class is found or the scan reaches the top of the inheritance hierarchy.

To illustrate this translation step, we need to extend our example and add some inheritance. We introduce a new class \ttt{Herring} (see \autoref{lst:carclassgroup}) that extends the \ttt{Fish} class.

\begin{lstlisting}[label=lst:carclassgroup,caption=Introducing the Herring class]
public class Herring extends Fish {
	@Group
	public void swarmAttack(Creature creature) {
		super.swarmAttack(creature);
		this.tauntVictim();
	}	
	@Group
	public void tauntVictim(){
		...
	}
}
\end{lstlisting}

This class is subject to the same translation steps as its parent class. This means an inner class that is responsible for \cb is generated and the methods are moved to the new class.  Furthermore, as the wrapper class of \ttt{Herring} should extend the wrapper class of \ttt{Fish}, that inheritance information is added to the tree. 

\subsection{Transformation of Swarm Behavior Call Sites}
As method resolution takes place during the Attribution phase, \JAG uses it as a pointcut for adding the previously generated wrapper classes to the AST.
If the original method lookup fails, we retry with the wrapper class of the call site, if such a class is present. If a \cm is found and hence the method call is valid we have to wrap that call. This results in passing the original call site as parameter to the wrapper class and let that wrapper class receive the original call.

This step concludes the syntax tree translations done by \JAG. The other passes of the compiler remain unchanged.

\section{Discussion}\label{discussion}\label{sec:discuss}

This section discusses issues related to both \CB in general as well as issues that are specific to the implementation of the \JAG prototype.

\paragraph{} ---
Currently, the only possible container type is \ttt{Collection}. This limits the interface available from within the \cm. The implementation must be expanded to enable any container type. For example if the programmer has to use the \ttt{java.util.Queue} interface he cannot invoke \ttt{Queue} specific methods in the current implementation. To solve this issue it must be possible to define the class or interface that is used as container for each \cm. The following line of code suggests a possible implementation that passes the container type as parameter to the annotation.
\begin{verbatim}
@Group("java.util.Queue")
\end{verbatim}

\paragraph{} ---
Within the lexical scope of a \cm defined in class \verb$T$ the pseudovariable \texttt{this} refers to a collection of \verb$T$ instances rather than to an instance of \verb$T$ as usual. This can lead to confusing situations when in the same file \ttt{this} refers to both an element instance in instance methods and to a container of elements in the \cms. For future releases of  \JAG, we thus plan to introduce a new pseudovariable \ttt{group} to address this issue.
 
\paragraph{} --- The synthetic wrapper classes use generics, this implies the limitation that from within a \cm it is not possible to create new instances of the container's element type. This is a general limitation of generics in Java.

\paragraph{} --- In previous work we presents a \CB in Smalltalk \cite{Kuhn07b}. Smalltalk is a dynamically typed language and thus  the \cb of a group is determined dynamically at runtime. The least common supertype of a collection changes as objects join or leave the group at runime. This results in a problem with empty containers, as the least common supertype of an emtpy container's elements is undefined. For the Smalltalk implementation we have chosen to fail with an exception when \cms are sendt to an empty collection. In \Java we can do better. We use the generic type paramater of collection instances to determine the element's type. This implies that the \cb of a group is not dynamic, as the least common supertype of its elements is declared at compile time. This solves the problem with emtpy collections mentioned above, since the generic type paramater is also known for empty instances.

\section{Related Work}\label{sec:rel} 
\label{sec:related}

\Cb is not the first approach to address group-related behavior. This section discusses related work on other group-related approaches.

There are also approaches such as \eg C++ generics, that allow the programmer to express \cb by seamlessly using the meta-programming facilities of the host languages. However, \cb targets a way more specific problem than general-purpose meta-programming approaches. \Cb extends the object model to decouple the behavior of collections from the collection hierarchy, and to extend collections with domain specific behavior that depends on the element type of a collection.

\subsection{Array Programming and Higher Order Messages}

\Cb is not the same as array programming (\ie projection of message sends). The principle behind array programming is that the same operation is applied to an entire array of data, 
without the need for explicit loops. Apart from ancient languages such as APL, or mathematical software such as MathLab and Mathematica, array programming has been recently applied in the context of dynamic \OO languages by FScript~\cite{Moug03a}, a Smalltalk-based scripting language for OSX, and by the ECMAScript for XML (E4X) specification, an extension of JavaScript.

Higher Order Messages (HOM)~\cite{Weih05a} can be seen as a generalization of array programming in the context of dynamic object-oriented languages. 
The basic operation in \OO languages is the message send. The fundamental principle behind HOM programming is that the same message is send to multiple objects. 
The principle of HOM programming is also known as cooperative call or message broadcast. The objects that a message is sent to can be those of an array or a collection, 
resulting in a form of array programming. In addition, HOM can be used on any composite structure of objects. It is not limited to arrays or collections.

Compared to swarm behavior, both array programming and higher order messages are too limiting to be useful in defining domain-specific behavior for collections of objects.
Their concern is to manipulation groups of objects at once, but not to attach custom behavior to the group as a whole. Consider for example \texttt{showTagCloud()},
its semantics can not be achieved with messaging alone. \Cb provides the required extension, as it allows the group's elements to attach custom behavior to the group.  

\subsection{Generics in Java and C\#}

This subsection compares \cb to Java generics, as sometimes, programmers tend to confuse these two concepts. Even though both generics and \cb are concerned with element types, they have not much in common. The concern of generics is type safety. A generic class \texttt{List<T>} is parameterized at compile-time, \eg as \texttt{List<Book>}, to ensure type-safety of its elements. Generics have recently been introduced in C\# and Java. 

The concern of \cb, on the other hand,  is to extend a collection with domain-specific behavior based on the type of the collection's elements. Consider again \texttt{showTagCloud()}, that is defined in the class \texttt{Book} as a \emph{\cm}. Instead of having to write a specific \texttt{BookList} class, using \cb, any collection who's elements are of type \texttt{Book} understands \texttt{showTagCloud()}, any collection of any collection type.

\subsection{Predicate types}

\Cb may be considered a special kind of predicate classes \cite{Cham93a}. Like a normal class, a predicate class has a superclass, methods and fields. However, unlike normal classes, any object may automatically become an instance of a predicate class whenever it satisfies a predicate condition associated with a predicate class. Considering again the library example, the \cb of \texttt{Book} is a predicate class \texttt{Book group} with the following predicate (given in OCL syntax)

\begin{alltt}\small
    context Book group pred:
    self isKindOf Collection and
        self->forAll( each | each isKindOf Book )    
\end{alltt}     

\subsection{Traits}
Traits \cite{Duca06b} are collections of methods (behavior) that can be composed into classes. Traits are normally seen as entities that are composed by the developer when designing the system. More dynamic notions of traits have been explored where traits are installed or retracted at runtime \cite{Berg05d}.
As traits are applied to classes, they provide behavior to all instances of the class, similar to normal methods. Thus traits alone do not help to model behavior of a collection 
of objects. It would be interesting to explore a combination of traits with \cb. Traits could be used to structure the \cb. As classes are
composed of traits, groups could be composed from traits to support reuse and limit code-duplication.

\subsection{Context Oriented Programming}
ContextL \cite{Cost05a, Hirs08a} is a language to support Context-Oriented Programming (COP).
The language provides a notion of \emph{layers}, which package context-dependent behavioural variations.
In practice, the variations consist of method definitions, mixins and \emph{before} and \emph{after} specifications.
Layers are dynamically enabled or disabled based on the current execution context. One can see \cb as a form of context orientation: a collection has, 
in the context that all the elements are of a certain type, different behavior then in other contexts. Current COP languages do not provide a model 
for context that is powerful enough to express \cb and therefore does not provide any help for modeling group behavior.

\section{Conclusion}\label{sec:final}\label{sec:fazit} 
	
This paper presents \emphCB, a new composition operator to associate behavior with a collection of instances. \Cb extends the object model to decouple the behavior of collections from the collection hierarchy. \Cb allows programmers to extend collections with domain specific behavior that depends on the element type of a collection. 

Most current languages offer limited support for group-related behavior. We report and analyze three idioms that are commonly used achieve group-related behavior. From this analyses we conclude a set of required properties for group-related behavior and propose \CB as a solution.

\begin{itemize}
\item \Cb is defined in the lexical scope of the element class.
\item \Cb is restricted to collections of the defining element type.
\item \Cb is not restricted to a specific subtype of collection. 
\item \Cb parallels the inheritance hierarchy of the element types.
\end{itemize}

We provide $\proc{Swarm-Lookup}$ as an algorithm for \cb lookup. \proc{Swarm-Lookup} extend the normal lookup of collections. If no instance method definition is provided by the collection's class, the most specific superclass of the collection elements is used as the base for a group lookup. Group lookup works in the same way as instance lookup, but operates on group methods rather than instance methods. Group methods are defined in the lexical scope of a class using the \verb$@Group$ annotation. Within a group method the pseudovariable \verb$this$ is bound to a collection of instances of the defining class.

We present the \JAG prototype, an implementation of \CB for Java. Our prototype implements \CB by extending the \Java compiler with a series of AST transformations. The prototype is open source and available online for download (see \autoref{sec:A}).

As future work we would like to use the \JAG compiler to refactor major Java applications towards \CB. Previous work using a Smalltalk implementation of \CB yielded promising preliminary results (upto 20\% code reduction \cite{Kuhn07b}) in this direction. Also we would like to further formalize the operational semantics of \CB.

\section*{Acknowledgments} 

We thank Oscar Nierstrasz for his feedback and suggestions. We thank Toon Verwaest for his support during the first steps of the \JAG project. 
We gratefully acknowledge the financial support of the Swiss National Science Foundation for the projects ``Bringing Models Closer to Code" (SNF Project No.\ 200020-121594, Oct.\ 2008 - Sept.\ 2010) and ``Biologically inspired Languages for Eternal Systems'' (SNF Project No.\ PBBEP2-125605, Apr.\ 2009 - Mar.\ 2010).

\bibliographystyle{plain}
\bibliography{scg}

\appendix

\section{\JAG Quickstart}\label{sec:A}
Our prototype implementation of \cb, \JAG, is available online at the following URL:

{\small \begin{lstlisting}
https://www.iam.unibe.ch/scg/svn_repos/Students/erni/bachelor/dist/jag.zip
\end{lstlisting}}

The archive consists of 2 Files, \ttt{classes.jar} that contains the modified compiler classes and \ttt{jagc} that wraps the default {javac} command and adds the parameters necessary to load the modified compiler classes. Note that you will need at least Java 6 to be able to run \JAG. Our modification enables \CB by loading the modified Java compiler classes into the bootclasspath and hence overrides the default classes. This can be done by either 
using the above mentioned \ttt{jagc} shell script (for unix systems) that automates everything:
\begin{verbatim}
jagc <*.java>
\end{verbatim}
or by adding the bootclasspath and classpath manually using the following command:
\begin{verbatim}
javac -J-Xbootclasspath/p:<path to classes.jar> 
	-cp <path to classes.jar> <*.java>
\end{verbatim}

It is necessary to add classes.jar to the classpath because the annotation \ttt{@Group} is provided by that file. 
The annotation must be imported from \ttt{com.sun.tools.javac.group.Group}.

The following classes serve as example and show the basic functionality of \JAG.  The \ttt{Creature}, \ttt{Fish} and \ttt{Shark} classes model the running example of this paper. The \ttt{ExampleFight} class is the main class of the example that prints out the result to the command line.

These classes can now be compiled by typing the command
\begin{verbatim}
./jagc Creature.java Fish.java ExampleFight.java Shark.java 
\end{verbatim}
Compilation results in 5 class files, Creature.class, Fish.class, Shark.class, ExampleFight.class and  Fish\$Group\$Fish.class that represents the inner class.
The compiled program can now be started by typing
\begin{verbatim}
java ExampleFight
\end{verbatim}
This prints the following on the command line:
\begin{verbatim}
Shark alive with 100 hitpoints.
Shark defeated with -9900 hitpoints.
\end{verbatim}
The source code:
\needlines{12}
\begin{lstlisting}
public class Creature {
	
	protected int hitpoints;

	public void damage(int damage) {
		this.hitpoints -= damage;	
	}
	
	public boolean alive() {
		return hitpoints > 0;
	}
}
\end{lstlisting}

\needlines{10}
\begin{lstlisting}
public class Shark extends Creature {
	
	public Shark() {
		this.hitpoints = 100;
	}
	
	public void attack(Creature creature) {
		creature.damage(10);
	}
}
\end{lstlisting}

\needlines{19}
\begin{lstlisting}
import com.sun.tools.javac.group.Group;

public class Fish extends Creature {
	
	public Fish() {
		this.hitpoints = 1;
	}
	
	public void attack(Creature creature) {
		creature.damage( 0 );
	}
	
	@Group
	public void swarmAttack(Creature creature) {
		int swarmSize = this.size();
		creature.damage( swarmSize * swarmSize );
	}	
	
}
\end{lstlisting}

\needlines{19}
\begin{lstlisting}
import java.util.ArrayList;
import java.util.Collection;

public class ExampleFight {

	public static void main(String[] args) {
		// setup
		Shark shark = new Shark();
		Collection<Fish> swarm = new ArrayList<Fish>();
		for (int n = 0; n < 100; n++) swarm.add(new Fish());
		
		// example 1, aggregation
		for (Fish fish: swarm) fish.attack(shark);
		System.out.printf("Shark alive with \%d hitpoints.\n", shark.hitpoints);
		// ==> 100
		
		// example 2, swarm behavior
		swarm.swarmAttack(shark);
		System.out.printf("Shark defeated with \%d hitpoints.\n", shark.hitpoints);
		// ==> 100 - (10000) = -9900 
	}
}
\end{lstlisting}

\end{document}